# Predicting the optimized thermoelectric performance of MgAgSb

C. Y. Sheng, H. J. Liu[*], D. D. Fan, L. Cheng, J. Zhang J. Wei, J. H. Liang, P. H. Jiang, J. Shi

*Key Laboratory of Artificial Micro- and Nano-Structures of Ministry of Education and School of Physics and Technology, Wuhan University, Wuhan 430072, China*

Using first-principles method and Boltzmann theory, we provide an accurate prediction of the electronic band structure and thermoelectric transport properties of $\alpha$-MgAgSb. Our calculations demonstrate that only when an appropriate exchange-correlation functional is chosen can we correctly reproduce the semiconducting nature of this compound. By fine tuning the carrier concentration, the thermoelectric performance of $\alpha$-MgAgSb can be significantly optimized, which exhibits a strong temperature dependence and gives a maximum *ZT* value of 1.7 at 550 K.

With the increasing energy demands, it is urgent to explore renewable and clean sources of energy. Thermoelectric materials have attracted much attention as they can directly convert heat into electricity and vice versa. The conversion efficiency of thermoelectric materials is determined by the dimensionless figure of merit:

$$ZT = \frac{S^2\sigma}{\kappa_e + \kappa_l}T, \qquad (1)$$

where $S$, $\sigma$, $T$, $\kappa_e$, $\kappa_l$ are the Seebeck coefficient, the electrical conductivity, the absolute temperature, the electronic and lattice thermal conductivity, respectively. A higher *ZT* value suggests a better thermoelectric performance, and one must therefore try to maximize the power factor ($PF = S^2\sigma$) and simultaneously minimize the thermal conductivity ($\kappa = \kappa_e + \kappa_l$). It is however very difficult to do so since these transport coefficients are usually coupled with each other. Over the past decades, $Bi_2Te_3$-based alloys [1, 2] are found to exhibit very good thermoelectric performance in the low temperature region. Unfortunately, their large-scale commercial applications are limited by the scarcity of Te element. It is therefore necessary to develop alternative high-performance thermoelectric materials containing elements relatively cheap and abundant in earth.

---

[*] Author to whom correspondence should be addressed. Electronic mail: phlhj@whu.edu.cn

Recently, MgAgSb compound has been proposed as a promising thermoelectric material due to its intrinsic low thermal conductivity and moderate electronic transport performance. Kirkham *et al.* [3] found that the compound exists in three different crystal structures, taking a tetragonal structure at room temperature ($\alpha$-MgAgSb), a $Cu_2Sb$-related structure at intermediate temperature ($\beta$-MgAgSb), and a half-Heusler structure at high temperature ($\gamma$-MgAgSb). Among them, the highest *PF* can be observed in the $\alpha$ phase. In the subsequent works, many efforts were made to improve the thermoelectric performance of $\alpha$-MgAgSb. For example, Zhao *et al.* [4] replaced Ag with Ni and obtained a maximum *ZT* of about 1.4. Shuai *et al.* [5] investigated the effect of Na doping in the Mg site and found that it is effective to improve *PF* but much higher *ZT* could not be achieved. By substitution of Ag with Cu [6], a *ZT* value of 1.32 can be reached at 250 °C for the $MgAg_{0.963}Cu_{0.007}Sb_{0.99}$ compound. In addition, Liu *et al.* fabricated the $MgAg_{0.97}Sb_x$ compound with different Sb content, and obtained a leg efficiency of 10.1% at the optimized content x = 0.995 [7]. On the theoretical side, Ying *et al.* [8] calculated the energy band structure of $\alpha$-MgAgSb using first-principles projector augmented wave (PAW) method [9, 10], and found it is semiconducting with an indirect gap of 0.1 eV. In contrast, calculations of Miao *et al.* [11] suggested that $\alpha$-MgAgSb is a semimetal by using the same approach. Further theoretical calculation is therefore desired to determine the real band structure of $\alpha$-MgAgSb, so that the transport and thermoelectric properties of this compound can be accurately predicted. In the present work, density functional calculations with different levels of exchange-correlation functional are performed to study the electronic properties of $\alpha$-MgAgSb. Our results indicate that the compound is semiconducting with an indirect band gap of 0.3 eV. Moreover, it is found that at optimized carrier concentration, a *ZT* value as high as 1.7 can be reached at 550 K.

Our theoretical calculations are performed within the framework of density functional theory (DFT), by using the PAW [9, 10] method as implemented in the Vienna *ab initio* simulation package (VASP). The exchange-correlation energy is in the form of Perdew-Bruke-Ernzerhof (PBE) [12] and the modified Becke-Johnson (mBJ) [13]. For comparison, we also consider the hybrid density functional in the form of Heyd-Scuseria-Ernzerhof (HSE) [14]. The cutoff energy for the plane-wave basis set is 500 eV, and a $10\times10\times10$ $\Gamma$ centered ***k***-mesh is used for the Brillouin zone integrations. The atom positions are fully relaxed until the magnitude of the force acting on each atom becomes less than 0.01 eV/Å. The thermoelectric transport coefficients are calculated by using the semiclassical Boltzmann theory [15] with relaxation time approximation.

As shown in Figure 1, $\alpha$-MgAgSb is a ternary compound with equal compositions 1:1:1 and crystallizes in a body-centered tetragonal structure having the space group $I\bar{4}c2$ (120). It can be viewed as a distorted Mg-Sb rocksalt lattice and half of the Mg-Sb pseudocubes are filled with Ag atoms. There are 48 atoms in the unit cell (Fig. 1 (a)), and 24 atoms in the primitive cell (Fig. 1 (b)). The calculated lattice constants are $a=b=9.152$ Å and $c=12.943$ Å, which are very close to that measured experimentally [3].

Figure 2 displays the electronic band structure of $\alpha$-MgAgSb along high symmetry lines in the irreducible Brillouin zone, where calculations using several different forms of exchange-correlation functional are shown for comparison. For the standard PBE functional, we see from Fig. 2 (a) that $\alpha$-MgAgSb is "semi-metallic". This is however questionable since it was experimentally demonstrated that $\alpha$-MgAgSb is semiconducting with an estimated band gap of 0.16 eV [4]. Moreover, first-principles calculations using the local density approximation (LDA) also found it is a semiconductor with a band gap of 0.1 eV [8]. As the standard DFT calculations usually underestimates the band gap seriously, we repeat the band structure calculation by using the mBJ functional, which is believed to yield band gaps with high accuracy since the Becke-Roussel potential in mBJ mimics the Coulomb potential created by the exchange hole [16]. As can be see from Fig. 2 (b), the compounds exhibits an indirect band gap of 0.3 eV with the valence band maximum (VBM) located at the X point and the conduction band minimum (CBM) around the Γ point. To further confirm the reliability of our calculation, the band structure of α-MgAgSb is recalculated by using more accurate HSE functional. As can be seen from Fig. 2(c), both the band gap and the band shape of mBJ calculation agree well with that of HSE result. Since the computational effort involved in the HSE approach is much larger (especially for the calculations of transport properties), we will exclusively use the mBJ functional in the following discussions.

By using the semiclassical Boltzmann theory and the relaxation time approximation, the transport coefficients of α-MgAgSb can be derived from the band structure. Such approach has been generally accepted to predict the optimal doping level of some known thermoelectric materials [17, 18, 19]. To get reliable results, a dense $20 \times 20 \times 20$ $k$-mesh is used in the irreducible Brillouin zone. Figure 3 (a) plots the calculated Seebeck coefficient ($S$) as a function of carrier concentration at room temperature. Around the Fermi level, the Seebeck coefficient of α-MgAgSb exhibits two obvious peaks, which is 439 $\mu$V/K for $p$-type carriers and 279 $\mu$V/K for $n$-type system. For the electrical conductivity $\sigma$, we should note that it can only be calculated with respect to the relaxation time $\tau$ in the Boltzmann transport theory. The accurate determination of $\tau$ is usually very complicated and depends on the detailed scattering mechanism. For simplicity, here the relaxation time is estimated by fitting the experimentally measured electrical conductivity [8]. As shown in Fig 3 (b), there is a sharp increase of $\sigma$ around the band edges, which is more pronounced for the $p$-type system. However, we should mention that the Seebeck coefficient $S$ and the electrical conductivity $\sigma$ exhibit opposite behavior with the variation of carrier concentration. As a consequence, one should fine tune the carrier concentration so that the power factor $S^2\sigma$ can be maximized. As for the electronic thermal conductivity $k_e$, we see from Fig. 3 (c) that it basically exhibits the same behavior as that of the electrical conductivity $\sigma$. This is reasonable since $k_e$ is derived from $\sigma$ by using the Wiedemann-Franz law $k_e = L\sigma T$, where the Lorenz number $L$ can be obtained by [20]:

$$L = \frac{k_e}{\sigma T} = \left(\frac{k_B}{e}\right)^2 \left[\frac{(r+7/2)F_{r+5/2}(\xi)}{(r+3/2)F_{r+1/2}(\xi)} - \left(\frac{(r+5/2)F_{r+3/2}(\xi)}{(r+3/2)F_{r+1/2}(\xi)}\right)^2\right], \quad (2)$$

Here $F_n(\xi)$ is a function of the reduced Fermi energy $\xi$ given by $F_n(\xi) = \int_0^\infty \frac{x^n dx}{e^{x-\xi}+1}$. In the whole carrier concentration range from $1\times10^{19}$ to $1\times10^{21}$ cm$^{-3}$, the calculated Lorenz number of $\alpha$-MgAgSb is $1.5\times10^{-8} \sim 2.2\times10^{-8}$ V$^2$K$^{-2}$. On the other hand, the lattice thermal conductivity $k_l$ of $\alpha$-MgAgSb is taken from the experiment measurement [8]. As shown in Fig. 3 (d), $k_l$ decreases exponentially with increasing temperature $T$ and can be fitted by:

$$k_l = 25.94\exp(-T/56.95) + 0.72. \quad (3)$$

In the temperature range from 300 to 550 K, the $k_l$ of $\alpha$-MgAgSb is found to decrease from 0.85 to 0.72 W/mK. Such ultrasmall thermal conductivity may be ascribed to the special crystal structure of $\alpha$-MgAgSb, such as the large unit cell and distorted lattice [6]. Moreover, it was demonstrated by Li *et al.* [21] that the local atomic disorders induced by the concurrent migration of Ag$^+$ and Mg$^{2+}$ ions can also result in a very small lattice thermal conductivity.

We now move to the investigation of the thermoelectric performance. Inserting the above discussed electronic and phononic transport coefficients into Equation (1), the ZT values of $\alpha$-MgAgSb can be obtained. The results are shown in Figure 4 as a function of temperature and carrier concentration. By fine tuning the carrier concentration, we see that the ZT values of $\alpha$-MgAgSb can be optimized and exhibit obvious peaks around the Fermi level. At room temperature, the ZT values are actually very small, which is only 0.1 for the *n*-type system and 0.6 for the *p*-type system. However, there is a strong temperature dependence of the ZT values, which increase monotonically with increased temperature. For example, the ZT value of $p$-type system can be significantly enhanced from 0.6 to 1.7 as the temperature increases from 300 to 550 K. Detailed analysis of the corresponding coefficients (Table I) indicates that the optimized power factor of $\alpha$-MgAgSb is comparable with those of many good thermoelectric materials such as bismuth antimony telluride, half-Heusler compound, and filled skutterudite [2, 22, 23]. Together with the extremely low thermal conductivity, it is reasonable to find fairly good thermoelectric performance of $\alpha$-MgAgSb. By fine tuning the hole concentration to a value around $7.73 \times 10^{19}$ cm$^{-3}$, a maximum ZT value of 1.7 can be achieved at 550 K, which exceeds those reported experimentally and suggests there is still room to further enhance the thermoelectric performance of this well-studied compound.

It is natural to ask how to realize the optimized carrier concentration mentioned above. This can be done by doping the $\alpha$-MgAgSb with appropriate elements and content. Essentially, any of the three different atomic sites in the compound could be substituted by other atoms. We have a $p$-type system if the doped atom has less valence electron than that of the site atoms. Otherwise it is an $n$-type system. Table II summarizes all the possible doping elements and optimal doping content for the $p$-type and $n$-type $\alpha$-MgAgSb at 550 K. It is interesting to note that some of them have been already adopted in the experimental work. For example, our theoretical calculation suggests that it is favorable to optimize the ZT value by substituting Ag with Ni at a content of 0.53%. This is consistent with the experimental result that $\alpha$-MgAgSb has a higher ZT value with a nominal formula of MgAg$_{0.965}$Ni$_{0.005}$Ag$_{0.99}$ [4]. The many other possible doping atoms listed in Table II offer a promising way to efficiently improve the thermoelectric performance of $\alpha$-MgAgSb, which deserves further experimental investigations and confirmations.

In summary, we demonstrate by accurate first-principles calculations that $\alpha$-MgAgSb is semiconducting with an indirect band gap of 0.3 eV. With the help of Boltzmann transport theory, a maximal $p$-type ZT value of 1.7 is suggested at optimized carrier concentration ($7.73 \times 10^{19}$ cm$^{-3}$) and operating temperature (550 K). More importantly, our theoretical work provides a simple map by which one can efficiently find the best doping atoms and optimal doping content to maximize the thermoelectric performance of $\alpha$-MgAgSb.

We thank financial support from the National Natural Science Foundation (Grant No. 11574236 and 51172167) and the "973 Program" of China (Grant No. 2013CB632502).

**Table I** Optimized *p*- and *n*-type *ZT* values of $\alpha$-MgAgSb compound at different temperature. The corresponding carrier concentration, the transport coefficients, and the Lorenz number are also indicated.

| T (K) | p/n (cm$^{-3}$) | S ($\mu$V/K) | $\sigma$ (10$^4$ S/m) | $S^2\sigma$ (mW/mK$^2$) | L (10$^{-8}$V$^2$/K$^2$) | $\kappa_e$ (W/mK) | $\kappa_L$ (W/mK) | ZT |
|---|---|---|---|---|---|---|---|---|
| 300 | 6.18×10$^{19}$ | 200 | 5.45 | 2.18 | 1.62 | 0.26 | 0.85 | 0.59 |
| 350 | 6.36×10$^{19}$ | 219 | 5.45 | 2.61 | 1.59 | 0.30 | 0.78 | 0.85 |
| 400 | 6.56×10$^{19}$ | 234 | 5.45 | 2.98 | 1.58 | 0.34 | 0.74 | 1.10 |
| 450 | 6.87×10$^{19}$ | 245 | 5.52 | 3.31 | 1.56 | 0.39 | 0.73 | 1.33 |
| 500 | 7.22×10$^{19}$ | 253 | 5.61 | 3.58 | 1.56 | 0.44 | 0.72 | 1.54 |
| 550 | 7.73×10$^{19}$ | 256 | 5.82 | 3.82 | 1.56 | 0.50 | 0.72 | 1.72 |
| 300 | −5.30×10$^{19}$ | −150 | 1.65 | 0.37 | 1.73 | 0.09 | 0.85 | 0.12 |
| 350 | −5.66×10$^{19}$ | −158 | 1.84 | 0.46 | 1.70 | 0.11 | 0.78 | 0.18 |
| 400 | −6.21×10$^{19}$ | −161 | 2.11 | 0.54 | 1.70 | 0.14 | 0.74 | 0.25 |
| 450 | −6.91×10$^{19}$ | −161 | 2.44 | 0.63 | 1.70 | 0.19 | 0.73 | 0.31 |
| 500 | −7.82×10$^{19}$ | −157 | 2.88 | 0.71 | 1.71 | 0.25 | 0.72 | 0.37 |
| 550 | −8.89×10$^{19}$ | −153 | 3.40 | 0.80 | 1.72 | 0.32 | 0.72 | 0.42 |

**Table II** Possible doping elements and optimal doping content of $\alpha$-MgAgSb compound, which can be selected to maximize the thermoelectric performance at 550K. Results for *p*-type and *n*-type systems are both shown. The cited references indicate those experimental works reported previously.

| sites | *p*-type element | content | *n*-type element | content |
|---|---|---|---|---|
| Mg | Li, Na [5], K, Rb | 0.0053 | B, Al, Ga, In | 0.0061 |
| Ag | Ni [4], Pd, Pt | 0.0053 | Zn, Cd, Hg | 0.0061 |
|  | Co, Rh, Ir | 0.0027 |  |  |
| Sb | Si, Ge, Sn, Pb | 0.0053 | S, Se, Te | 0.0061 |
|  | Al, Ga, In [8], Tl | 0.0027 | Cl, Br, I | 0.0031 |

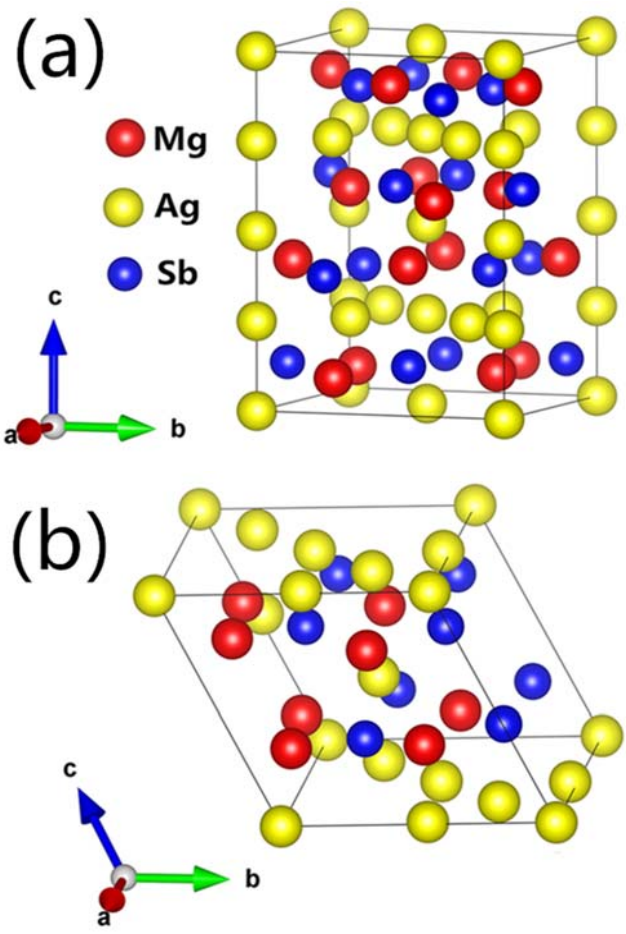

**Figure 1** The crystal structure of $\alpha$-MgAgSb: (a) unit cell, and (b) primitive cell.

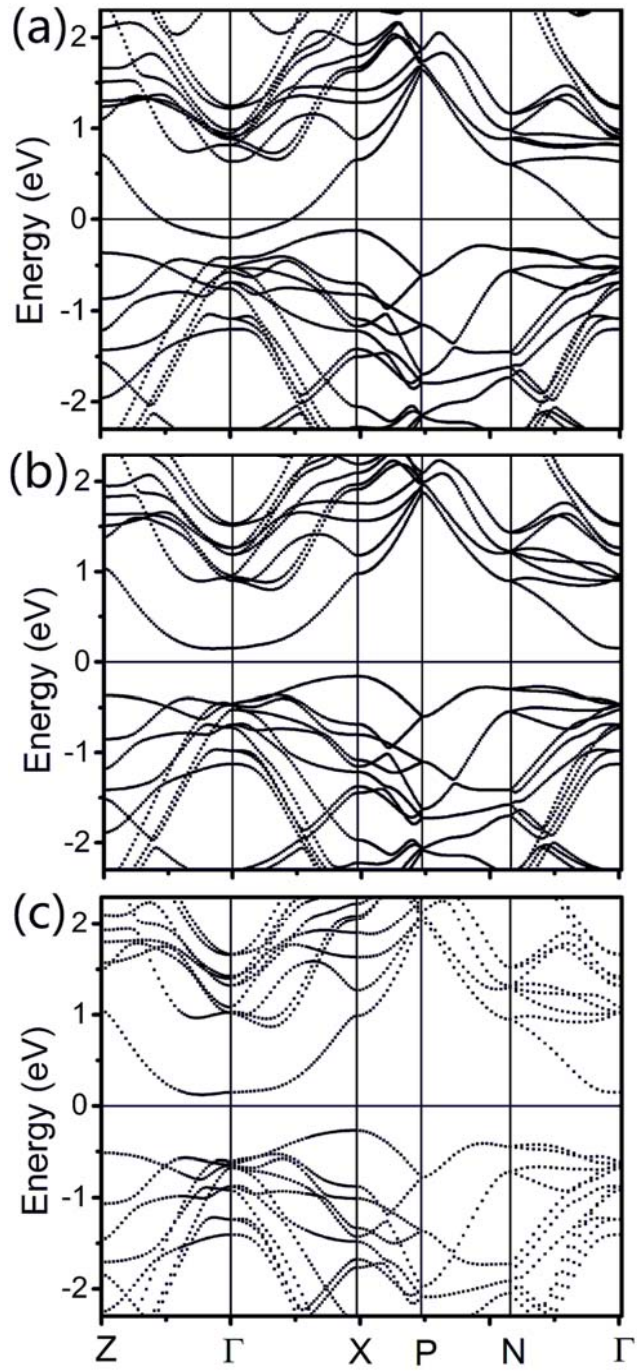

**Figure 2** Calculated band structure of α-MgAgSb by using (a) PBE, (b) mBJ, and (c) HSE functional.

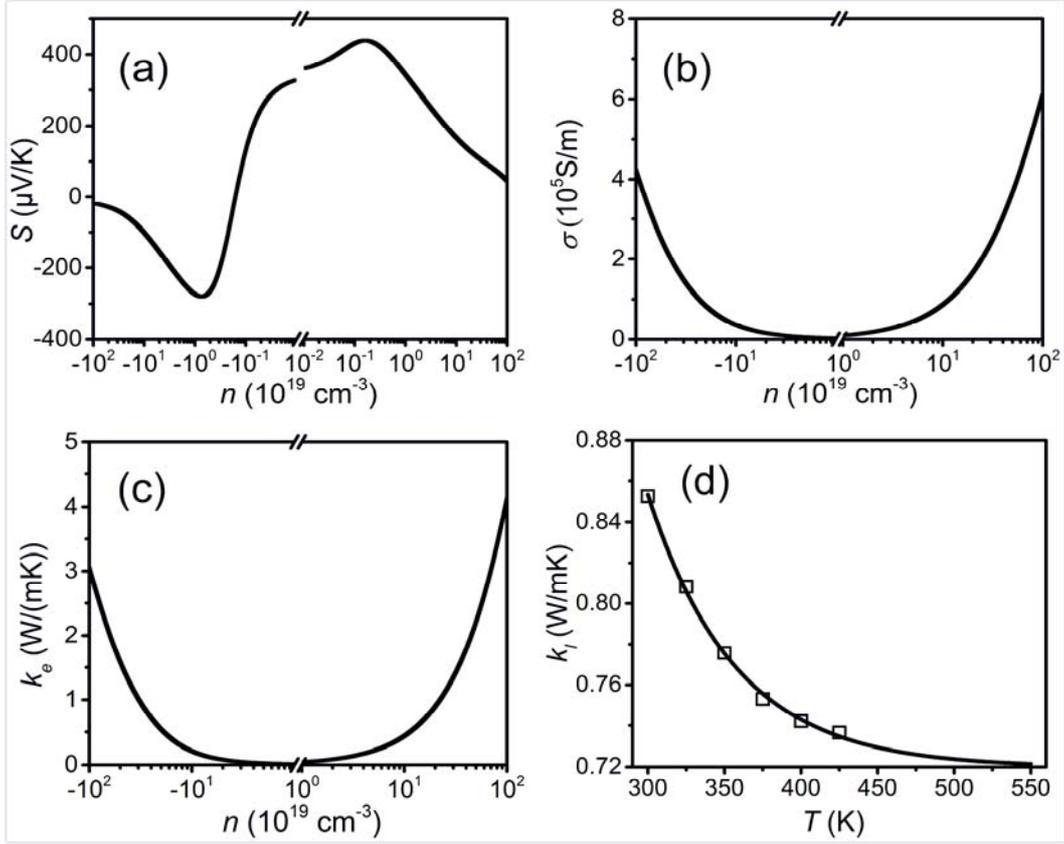

**Figure 3** Calculated (a) Seebeck coefficient, (b) electrical conductivity, and (c) electronic thermal conductivity of $\alpha$-MgAgSb as a function of carrier concentration at 300 K. Positive and negative carrier concentrations represent $p$- and $n$-type carriers, respectively. The lattice thermal conductivity of the system is shown in (d) as a function of temperature, which is obtained by fitting (line) the experimental results (squares) of Reference [8].

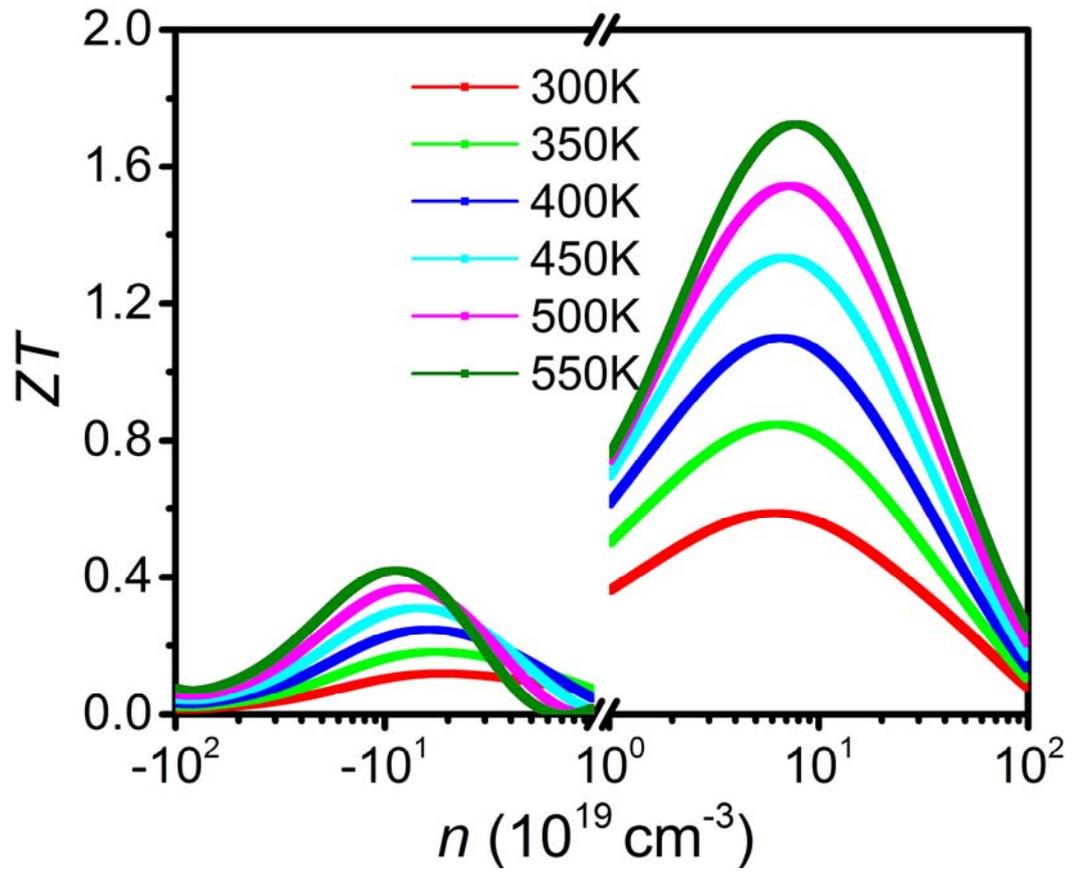

**Figure 4** Calculated ZT value of $\alpha$-MgAgSb as a function of temperature and carrier concentration.